\documentclass{article}
\makeatletter
\usepackage{setspace}
\usepackage{rotating}
\usepackage{multirow}
\usepackage{mathtools}
\usepackage{algorithm} 
\usepackage{tikz}
\usepackage{longtable}
\usetikzlibrary{shapes.geometric, arrows}

\tikzstyle{startstop} = [rectangle, rounded corners, 
text centered, 
draw=black]

\tikzstyle{io} = [trapezium, 
trapezium stretches=true, 
trapezium left angle=70, 
trapezium right angle=110,text centered, 
draw=black,text width=3cm,]

\tikzstyle{process} = [rectangle,  
draw=black,text width=1.5cm, text centered]

\tikzstyle{decision} = [diamond, 
draw=black,text width=1.5cm, text centered]
\tikzstyle{arrow} = [thick,->,>=stealth]
\usepackage{epsfig}
\usepackage{rotating}
\usepackage{amssymb}
\usepackage{xcolor}
\usepackage[T1]{fontenc}
\usepackage{lmodern}
\usepackage[utf8]{inputenc}

\usepackage{booktabs}
\usepackage{multirow}
\usepackage{enumerate}
\usepackage{caption}
\usepackage{subcaption}
\usepackage{lscape,float}
\usepackage{graphicx}
\usepackage{xcolor}
\usepackage{soul}
\usepackage[toc]{appendix}

\usepackage{natbib,hyperref,doi}
\makeatletter
\renewcommand\@biblabel[1]{#1.}
\makeatother
\usepackage[noend]{algpseudocode}
\usepackage{enumitem}

\usepackage{amsmath}
\usepackage{enumitem}

\makeatletter
\renewcommand\@biblabel[1]{#1.}
\makeatother

\usepackage{float,lscape}
 \providecommand{\keywords}[1] {
  \textit{Keywords:} #1}
\title{Determination of Optimum Warranty Region for Two Dimensional Dependent Data}

\author{
    Rathin Das$^1$ \and
    Tanmay Sen$^1$\thanks{Corresponding author\\Email address: tanmay.sen@isical.ac.in} \and
    Ritwik Bhattacharya$^2$ \and
    Biswabrata Pradhan$^1$
}
\date{$^1$ SQC \& OR Unit, Indian Statistical Institute, Kolkata, India\\
$^2$ Department of Mathematical Sciences, University of Texas El Paso, El Paso, TX}
\usepackage{siunitx}

\begin{document}

\maketitle

\begin{abstract}
This paper presents a method for determining the optimal two-dimensional warranty region for age and mileage scales across all possible combinations of free replacement warranty (FRW), prorata warranty (PRW), and FRW-PRW combined policies. The operational time or lifetime and usage of the products are modeled using a bivariate Gumbel copula with Weibull as the marginal distribution. The optimal warranty region is derived by maximizing an expected utility function, which incorporates two cost components: the economic benefit function and the warranty cost function, specifically constructed for the two-dimensional warranty scenario. To obtain the optimal warranty region, a real-world dataset of traction motors, including age and mileage information, is analyzed. The results show that considering a two-dimensional warranty cost function yields the highest utility compared to all other scenarios.

\end{abstract}


\section{Introduction}



In today’s competitive market, warranty agreements are standard for almost all products, driven by industry regulations, customer expectations, and business competition. These service contracts are important because they provide customers with post-purchase protection, offering solutions like repairs, replacements, or refunds if a product fails or doesn’t work as expected during the warranty period. Manufacturers often use different warranty policies as a marketing tool to increase sales, while customers consider the warranty as a key factor when choosing between similar products. As a result, warranty analysis is essential to improving the quality of manufactured products. If not done properly, it can harm business goals and damage how customers perceive the product. Manufacturers offer warranties to build trust in their product’s quality. However, offering an overly long warranty can lead to high costs for the manufacturer, while a shorter warranty compared to competitors can hurt sales. Therefore, design of optimal warranty period / region is a crucial task to the manufacturer. This is typically done through reliability assessments, often using life-testing experiments. 

The most commonly used warranty policies are the free replacement warranty (FRW) policy (see \cite{blischke_2006}), pro-rata warranty (PRW) policy (see \cite{menke1969determination}), and the combined FRW-PRW policy (see \cite{thomas1983optimum}). A key feature of these policies is that if a product fails during the warranty period, the consumer will receive either full or pro-rated compensation from the manufacturer. Under the FRW policy, if a product fails, a non-repairable product is replaced with an identical one at no cost. For repairable products, the manufacturer will repair the product free of charge. Under the PRW policy, however, the manufacturer provides pro-rated compensation to the consumer if the product fails. In some cases, a combination of both policies is applied, which is referred to as the combined FRW-PRW policy. 


The warranty can be classified into one-dimensional and two-dimensional based on the number of variables involved. A one-dimensional warranty policy involves a single variable, such as the age or usage of the product, while a two-dimensional warranty policy covers two variables, typically age and usage. Numerous studies focus on determining one dimensional warranty lengths for various policies either using complete or censored data. In the case of censored data, \cite{wu2007optimal} developed a cost model to determine the optimal burn-in time and warranty length for non-repairable products under the FRW-PRW policy within a classical framework. \cite{Christen_2006} determined the Bayesian optimal warranty length under a pro-rata warranty policy, using a two-parameter Weibull distribution to model product lifetime. \cite{wu_2010} explored a decision problem under the FRW-PRW policy, employing a Bayesian approach to establish optimal warranty lengths under a Type-II progressive censoring scheme with a Rayleigh distribution. \cite{sen2022determination} focused on determining the optimal warranty length for an FRW-PRW policy by deriving non-linear pro-rated rebate cost under Type-II Unified Hybrid Censoring schemes (Type-II UHCs).

Traditionally, warranties are determined by a single factor, such as time or usage. However, accounting for both factors provides a more accurate reflection of a product's wear and tear. Consequently, a two-dimensional warranty approach allows for a more precise calculation of the optimal warranty region and cost, benefiting both customer satisfaction and the manufacturer’s profitability. This type of warranty helps balance the interests of both the manufacturer and the consumer. From the manufacturer's perspective, it reduces the risk of excessive warranty claims while offering greater flexibility in pricing and coverage. For consumers, it ensures the product will be covered under more realistic conditions that reflect actual usage. In a two-dimensional warranty model, coverage is based not only on the product’s age but also on how much it has been used. This allows manufacturers to account for the fact that a product may wear out more quickly with higher usage, even if it’s relatively new, or conversely, it may last longer if used less frequently. For instance, a vehicle's warranty might cover it for a certain number of years or miles, whichever comes first. By considering both time and usage, manufacturers can design warranties that better align with the actual performance and failure patterns of their products.  

Products like automobiles, power generators, traction motors, and factory equipment are typically sold with a two-dimensional warranty, where both age and usage are considered together to determine warranty claim eligibility. \cite{lawless1995methods} investigated warranty claims for automobiles, considering the impact of both age and mileage accumulation on product reliability, and proposed a family of models that links failure to these two factors. \cite{jung2007analysis} suggests a method for estimating the parameters of a bivariate distribution that captures the positive correlation between age and usage for two dimensional warranty claim. \cite{gupta2014warranty} classified warranty regions based on manufacturing defects, usage, or fatigue, and proposed an approach with incomplete data to optimally differentiate these regions by estimating the change point in the hazard rate function. \citet{manna2006optimal,manna2007use,manna2008note} derived the two-dimensional expected warranty cost under the FRW warranty policy and determined a procedure for finding the optimal warranty region using the consumer's perspective.  \cite{yang2016warranty} proposed two models using a generalized renewal process to predict warranty claims and introduced the accelerated failure time (AFT) model to study the effect of usage rate on system reliability. Based on two-dimensional warranty data, \cite{dai2017field} employ an AFT model to examine how the usage rate affects product degradation, with the method being validated using real-world warranty data from an automobile manufacturer in China. 

This study considers determination of optimal warranty region by maximizing an expected utility function, which is composed of both economic benefit and warranty cost functions, for a combined FRW-PRW policy. The analysis is based on data obtained from a complete life testing scenario. To model the positively correlated two-dimensional data, we fit a bivariate Gumbel copula. The Weibull distribution is used to model both the age and usage scales. To the best of our knowledge, this study presents the first comprehensive approach for designing a two-dimensional warranty region, addressing all possible scenarios under complete sample. The main contributions of the proposed work are as follows:
\begin{enumerate}
    \item We propose a two-dimensional warranty cost function for all possible scenarios under a combined FRW-PRW policy.
    \item Economic benefit function is derived under combined policy. 
    \item We consider optimal design of warranty region by maximizing the expected utility.
    \item For illustration purpose we introduce the bivariate Gumbel copula to model age and usages together under two-dimensional warranty, when the marginal distributions are Weibull. 
\end{enumerate}

The rest of this article is organized as follows. In Section 2, two-dimensional positively correlated age and usage data is modeled by the Gumbel copula. In Section 3, all possible combinations of two-dimensional warranty policies associated with the cost of reimbursing time with age and usage are discussed. In Section 4, the economic benefit function and warranty cost function are discussed. In Section 5, determination of optimal warranty region is discussed. In Section 6, real-life data is used to illustrate the model. In Section 7, the conclusions are made.

\section{ Copula Based Life Time Models}
Copula was first introduced by \cite{sklar1959fonctions}, to model the dependence structure between random variables in multivariate distributions. Assume the two random variables $T$ for age and $U$ for usage follow Weibull distribution with cumulative distribution functions (cdf)
\begin{align*}
    F_{T}(t)=1-\exp\left[-\left(\frac{t}{\eta_T}\right)^{\lambda_T}\right]  ~\text{and}~  F_{U}(u)=1-\exp\left[-\left(\frac{u}{\eta_U}\right)^{\lambda_U}\right], ~\text{respectively.}
\end{align*}
where $\lambda_T$, $\lambda_U$ are scale parameters and $\eta_T, \eta_U$ are shape parameters of the marginal distributions. 
According to Sklar's theorem, the joint distribution function $F$ of $T$ and $U$ can be expressed in terms of its marginal distributions $F_T$ and $F_U$ and a copula function $C(\cdot,\cdot)$ such that 
\begin{align*}
F(t,u)=P(T\leq t,U\leq u)=C(F_{T}(t),F_{U}(u))~~t\geq 0,~u\geq 0.
\end{align*}
Likewise, the joint reliability function $R$ of $T$ and $U$ can be expressed in terms of its marginal reliability function $R_T$ and $R_U$ and a coupla function $\hat{C}(\cdot,\cdot)$ such that
\begin{align*}
R(t,u)=P(T>t,U>u)=\hat{C}(R_{T}(t),R_{U}(u))~~t\geq 0,~u\geq 0.
\end{align*}

\noindent We consider the Gumbel copula (\cite{nelsen2006introduction}), which belongs to the Archimedean copula family. Then joint cdf and the joint reliability function of $T$ and $U$ can be expressed as

\small\begin{eqnarray*}
    F(t,u)&=&C(F_T(t),F_U(u))\\
    &=&\exp\left\{-\left[\left(-\ln F_T(t)\right)^\theta+\left(-\ln F_U(u)\right)^\theta\right]^{1/\theta}\right\}\\
    &=&\exp\left\{-\left[\left(-\ln \left(1-\exp\left(-\left(\frac{t}{\eta_T}\right)^{\lambda_T}\right)\right)\right)^\theta+\left(-\ln\left( 1-\exp\left(-\left(\frac{u}{\eta_U}\right)^{\lambda_U}\right)\right)\right)^\theta\right]^{1/\theta}\right\},~\theta\geq 1 
\end{eqnarray*}
\noindent  
\begin{flalign*}
    R(t,u)& =\hat{C}(R_T(t),R_U(u))&& \\
    &= \exp\left\{-\left[\left(-\ln R_T(t)\right)^\theta+\left(-\ln R_U(u)\right)^\theta\right]^{1/\theta}\right\}
     = \exp\left[-\left\{\left(\frac{t}{\eta_{T}}\right)^{\lambda_T\theta}+\left(\frac{u}{\eta_{U}}\right)^{\lambda_U\theta}\right\}^{1/\theta}\right], ~\text{respectively}.
\end{flalign*}
\normalsize

\noindent The joint density function of $(T,U)$ can be written as
\begin{align*}
    f(t, u) = \frac{\partial^2 R(t,u)}{\partial t\partial u}
    =&\frac{\lambda_T\lambda_U}{\eta_T\eta_U}\left(\frac{t}{\eta_T}\right)^{\lambda_T\theta-1}\left(\frac{u}{\eta_U}\right)^{\lambda_U\theta-1}\left\{\left(\frac{t}{\eta_{T}}\right)^{\lambda_T\theta}+\left(\frac{u}{\eta_{T}}\right)^{\lambda_U\theta}\right\}^{1/\theta-2}\\
   & \times\left[ \left\{\left(\frac{t}{\eta_{T}}\right)^{\lambda_U\theta}+\left(\frac{u}{\eta_{U}}\right)^{\lambda_U\theta}\right\}^{1/\theta}+\theta-1\right]\exp\left[-\left\{\left(\frac{t}{\eta_T}\right)^{\lambda_T\theta}+\left(\frac{u}{\eta_{U}}\right)^{\lambda_U\theta}\right\}^{1/\theta}\right].
\end{align*}

\noindent Assume that \(n\) independent units are subjected to a life test. When an item fails, we observe two measurable quantities: age and usage. We assume that both the scale age and usage follow Weibull distributions, with parameters \((\eta_T, \lambda_T)\)  and \((\eta_U, \lambda_U)\), respectively. The joint distribution of age and usage is modeled using a Gumbel copula to capture their dependence. Under the complete sample scenario, the likelihood function for the parameter vector \(\boldsymbol{\psi} = (\eta_T, \lambda_T, \eta_U, \lambda_U, \theta)\), based on the observed sample \(\boldsymbol{D} = \{(t_1, u_1), \dots, (t_n, u_n)\}\), is expressed as:
\begin{eqnarray*}
    L(\boldsymbol{\psi}\ | \ \boldsymbol{D})=\prod_{i=1}^nf(t_i,u_i).
\end{eqnarray*}
The log-likelihood, derived under the complete sample, is given below  
\allowdisplaybreaks\begin{align}\label{log_like}
    l(\boldsymbol{\psi}\ | \ \boldsymbol{D})=&\log L(\boldsymbol{\psi}\ | \ \boldsymbol{D})\nonumber\\
    =&\sum_{i=1}^n\left\{\log(\lambda_T\lambda_U)-\log(\eta_T\eta_U)+(\lambda_T\theta-1)[\log(t_i)-\log(\eta_T)]+(\lambda_U\theta-1)[\log(u_i)-\log(\eta_U)]\right.\nonumber\\
    &+\left.\left(\frac{1}{\theta}-2\right)\log\left[\left(\frac{t_i}{\eta_{T}}\right)^{\lambda_T\theta}+\left(\frac{u_i}{\eta_{T}}\right)^{\lambda_U\theta}\right]+\log\left[ \left\{\left(\frac{t_i}{\eta_{T}}\right)^{\lambda_U\theta}+\left(\frac{u_i}{\eta_{U}}\right)^{\lambda_U\theta}\right\}^{1/\theta}+\theta-1\right]\right.\nonumber\\
    &-\left.\left[\left(\frac{t_i}{\eta_T}\right)^{\lambda_T\theta}+\left(\frac{u_i}{\eta_{U}}\right)^{\lambda_U\theta}\right]^{1/\theta}\right\}.
\end{align}

\section{Possible combination of two-dimensional warranty policies}
There are three types of one-dimensional warranty policies: free replacement warranty (FRW) policy, pro-rata warranty (PRW) policy, and combined FRW-PRW (CW) policy. First, we derive the cost functions for the three given policies based on a single-dimensional age/mileage scale. Next, we extend this approach to derive the cost functions for the two-dimensional case. 
Let $X$ be the random variable of a measurable quantity (ex. age, mileage, usage) to failure of a product. Consider the warranty in the period $(0,x_{w})$, $x_w \in \mathcal{R}$. 
Under the FRW policy, if a product fails within the warranty period, the manufacturer offers the consumer either a full refund or a free replacement with an identical product.  Let $S$ be the sales price of the product.  The cost of reimbursing items with $x$ is
\begin{align*}
    C_{FRW}^X(x)=S,& \text{ if } 0\leq x\leq x_w,
\end{align*}
where $x$ is the observed value of $X$. Under PRW policy, if the product fails during the given warranty period $(0,x_w)$, consumer receives a pro-rated compensation from the product manufacturer. The cost of providing prorated compensation is a linearly decreasing function of remaining useful age or mileage. The cost of reimbursing items with age/mileage $x$ is
\begin{align*}
    C_{PRW}^X(x)=S\left(1-\frac{x}{x_w}\right),& \text{ if } 0\leq x\leq x_w.
\end{align*}
Consider the CW policy, which is a combination of FRW and PRW policies. Under this CW policy, let us consider the warranty is given in the period  $(0,x_{w_2}]$ and also consider another  point $x_{w_1} \geq 0$ with $x_{w_1}\leq x_{w_2}$ such that the FRW is used in $(0,x_{w_1}]$ and PRW is used in $(x_{w_1},x_{w_2}]$. The cost of reimbursing items with $x$ is
\begin{align*}
    C^X_{CW}(x)=\begin{dcases}
        S &\text{ if } 0\leq x\leq x_{w_1}\\
        S\left(\frac{x_{w_2}-x}{x_{w_2}-x_{w_1}}\right)&\text{ if } x_{w_1}\leq x\leq x_{w_2}.\\
    \end{dcases}
\end{align*}

Different shapes of the warranty region are discussed by \cite{wang2018two} and \cite{blischke1992product} under combined two-dimentional warranty policy. 
To the best of our knowledge, the warranty cost function of reimbursing items based on the two-dimentional warranty policy is not defined yet. Here we have proposed the cost function under two-dimentional warranty policy by considering the shape of rectangular warranty region.
Let \( T \) be the random variable representing the age at failure of the product, and \( U \) be the random variable representing the usage at failure. A common example of a two-dimensional warranty policy is the rectangular-shaped warranty region offered by automobile manufacturers for cars, based on it's time to failure (age) and the number of tire revolutions (usage). Thus the warranty region can be derived as \( [0, t_w] \times [0, u_w] \), where \( t_w \) is the warranty period for age and \( u_w \) is the warranty period for usage. If the product fails within this region, the manufacturer provides warranty coverage. Consider the set of warranty policies $\mathcal{P} = \{FRW, ~PRW, ~CW\}$ for the scale age/mileage.  Thus the two dimensional warranty policy is composed of total nine $(3 \times 3 )$ scenarios such as $\{FRW \times FRW,~ FRW \times PRW, ~FRW \times CW, ~PRW \times FRW, ~PRW \times PRW, ~PRW \times CW, ~CW \times FRW, ~CW \times PRW, ~CW \times CW\}$. For $P_1,~P_2\in \mathcal{P}$, the cost of reimbursing an item with age $t$ and usage $u$ is defined as
\begin{align}\label{warranty}
    C_{P_1\times P_2}^{T\times U}(t,u)=\frac{1}{S}C_{P_1}^T(t)\times C_{P_2}^U(u).
\end{align}

To construct the two-dimensional warranty cost function for reimbursing an item, the individual cost functions for reimbursement based on the individual scale age and usage must first be known. Using the formula provided in (\ref{warranty}), the combined cost function for a two-dimensional warranty in a rectangular region can be derived.  All nine types of two-dimensional warranty policies associated with the warranty cost function of reimbursing an item with age $t$ and mileage $u$ are provided as follows. The corresponding graphical plot of reimbursement for an item with age $t$ and mileage $u$ is shown in Figures 1-9.

\begin{enumerate}[label=Case-\Roman*. , left=0pt] 
    \item [Case-I: ] $FRW\times FRW$
\begin{align*}
    C_{FRW\times FRW}^{T\times U}(t,u)=S, & \text { if } 0\leq t\leq t_w ~~0\leq u\leq u_w
\end{align*}
\begin{figure}[hbt!]
    \centering
  \includegraphics[scale=0.25]{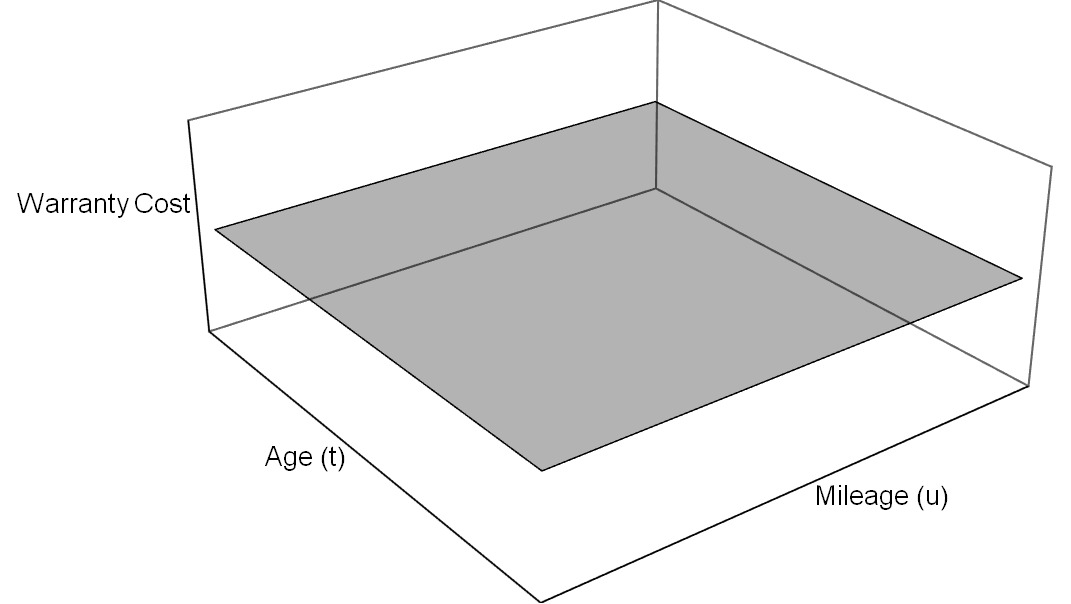}
    \caption{Warranty cost with age $t$ and mileage $u$ corresponding to case-I}
    \label{Case-I}
\end{figure}

    \item [Case-II: ]  $FRW\times PRW$
\begin{align*}
    C_{FRW\times PRW}^{T\times U}(t,u)=S\left(1-\frac{u}{u_w}\right), & \text { if } 0\leq t\leq t_w, ~~0\leq u\leq u_{w}
\end{align*}
\begin{figure}[hbt!]
    \centering
    \includegraphics[scale=0.25]{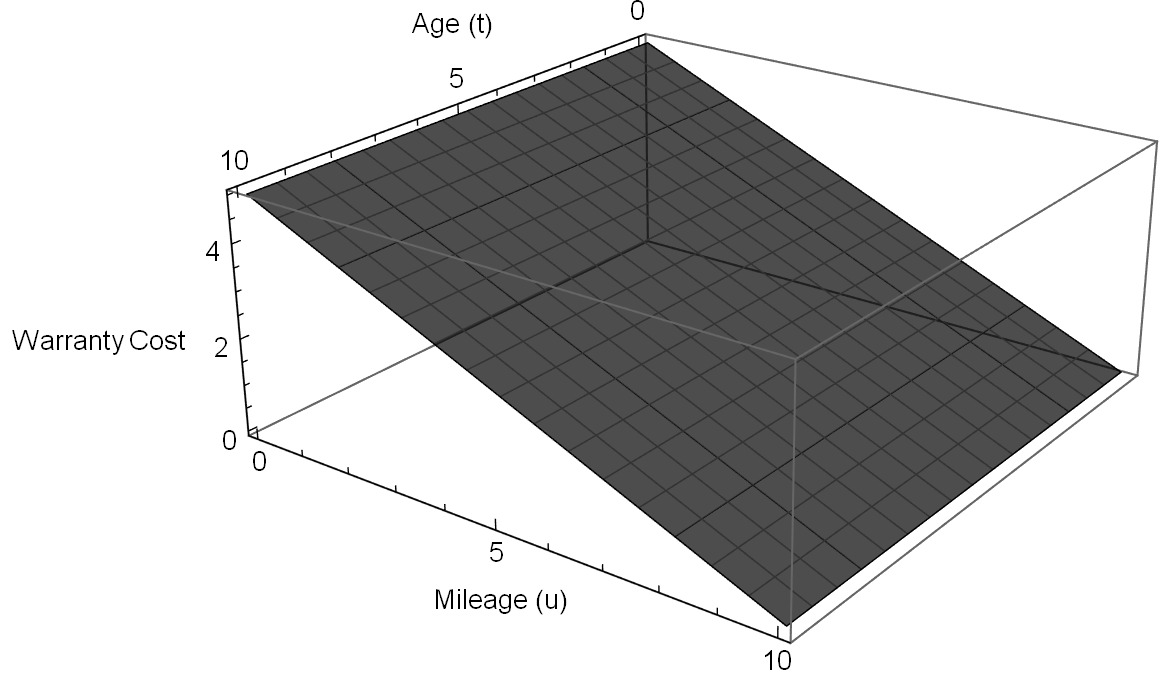}
    \caption{Warranty cost with age $t$ and mileage $u$ corresponding to case-II}
    \label{Case-II}
\end{figure}

    \item [Case-III: ] $FRW\times CW$
\begin{align*}
    C_{FRW\times CW}^{T\times U}(t,u)=\begin{dcases}S & \text { if } 0\leq t\leq t_w, ~~0\leq u\leq u_{w_1}\\
    S\left(\frac{u_{w_2}-u}{u_{w_2}-u_{w_1}}\right)&\text{ if } 0\leq t\leq t_w, ~~u_{w_1}\leq u\leq u_{w_2}
    \end{dcases}
\end{align*}
\begin{figure}[hbt!]
    \centering
    \includegraphics[scale=0.3]{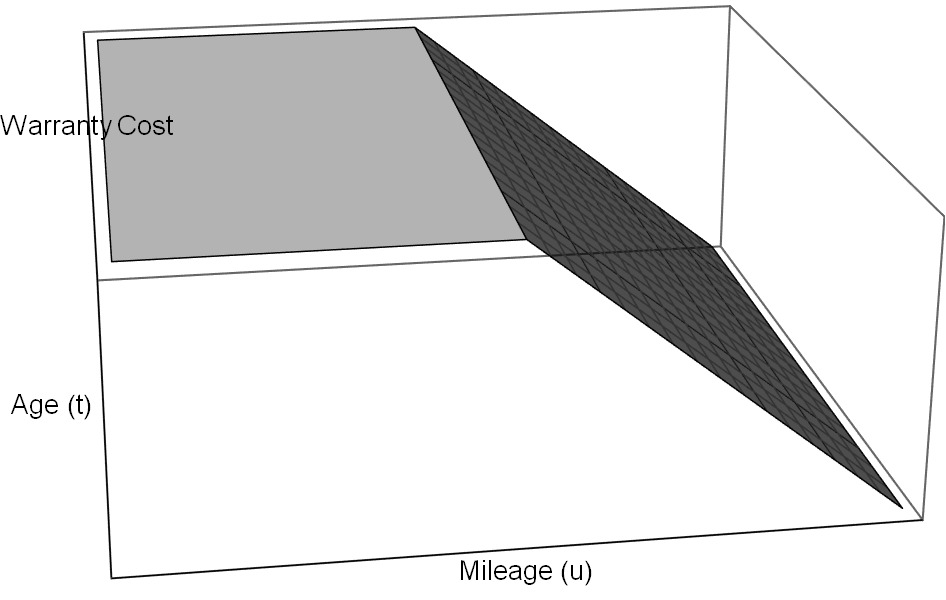}
    \caption{Warranty cost with age $t$ and mileage $u$ corresponding to case-III}
    \label{Case-III}
\end{figure}
    \item [Case- IV: ] $PRW\times FRW$
\begin{align*}
    C_{PRW\times FRW}^{T\times U}(t,u)=S\left(1-\frac{t}{t_w}\right), & \text { if } 0\leq t\leq t_w, ~~0\leq u\leq u_{w}
\end{align*}
\begin{figure}[hbt!]
    \centering
    \includegraphics[scale=0.3]{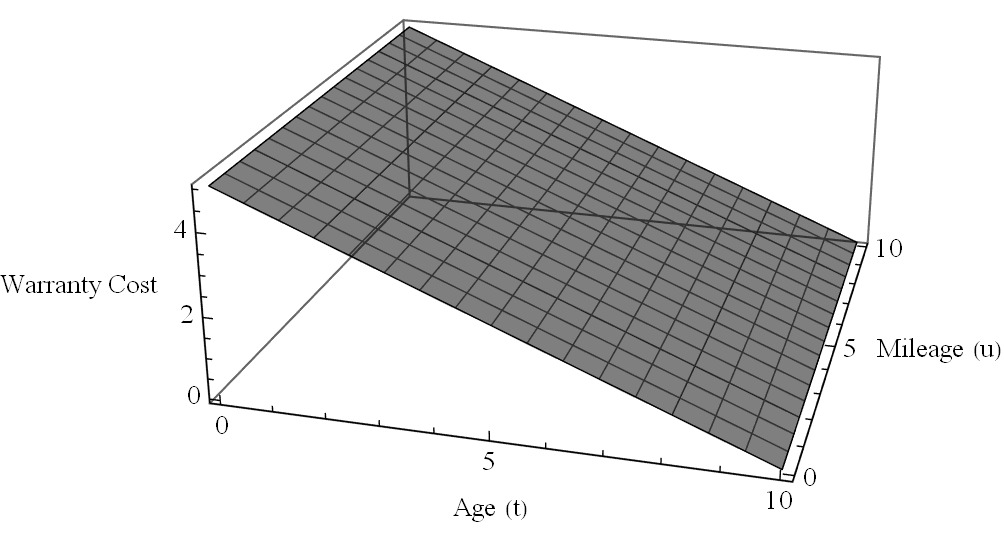}
    \caption{Warranty cost with age $t$ and mileage $u$ corresponding to case-IV}
    \label{Case-IV}
\end{figure}
    \item [Case-V: ] $PRW\times PRW$
\begin{align*}
    C_{PRW\times PRW}^{T\times U}(t,u)=S\left(1-\frac{t}{t_w}\right)\left(1-\frac{u}{u_w}\right),& \text { if } 0\leq t\leq t_w, ~~0\leq u\leq u_{w}
\end{align*}
\begin{figure}[hbt!]
    \centering
    \includegraphics[scale=0.3]{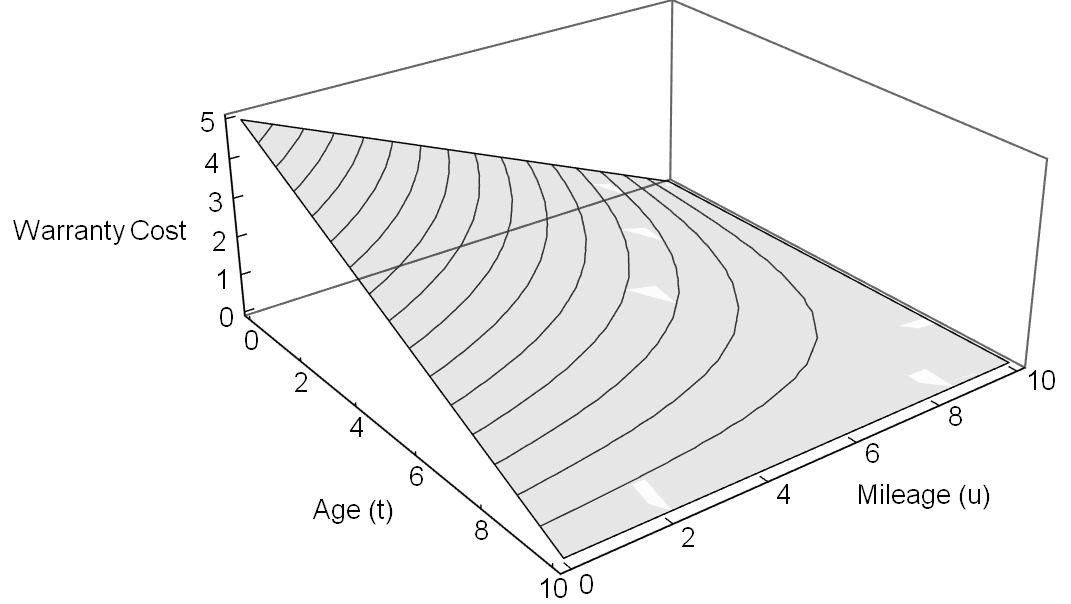}
    \caption{Warranty cost with age $t$ and mileage $u$ corresponding to case-V}
    \label{Case-V}
\end{figure}
    \item [Case-VI: ] $PRW\times CW$
\begin{align*}
    C_{PRW\times CW}^{T\times U}(t,u)=\begin{dcases}S\left(1-\frac{t}{t_w}\right)& \text { if } 0\leq t\leq t_w, ~~0\leq u\leq u_{w_1}\\
 S\left(1-\frac{t}{t_w}\right)   \left(\frac{u_{w_2}-u}{u_{w_2}-u_{w_1}}\right)&\text{ if } 0\leq t\leq t_w, ~~u_{w_1}\leq u\leq u_{w_2}
    \end{dcases}
\end{align*}
\begin{figure}[hbt!]
    \centering
    \includegraphics[scale=0.25]{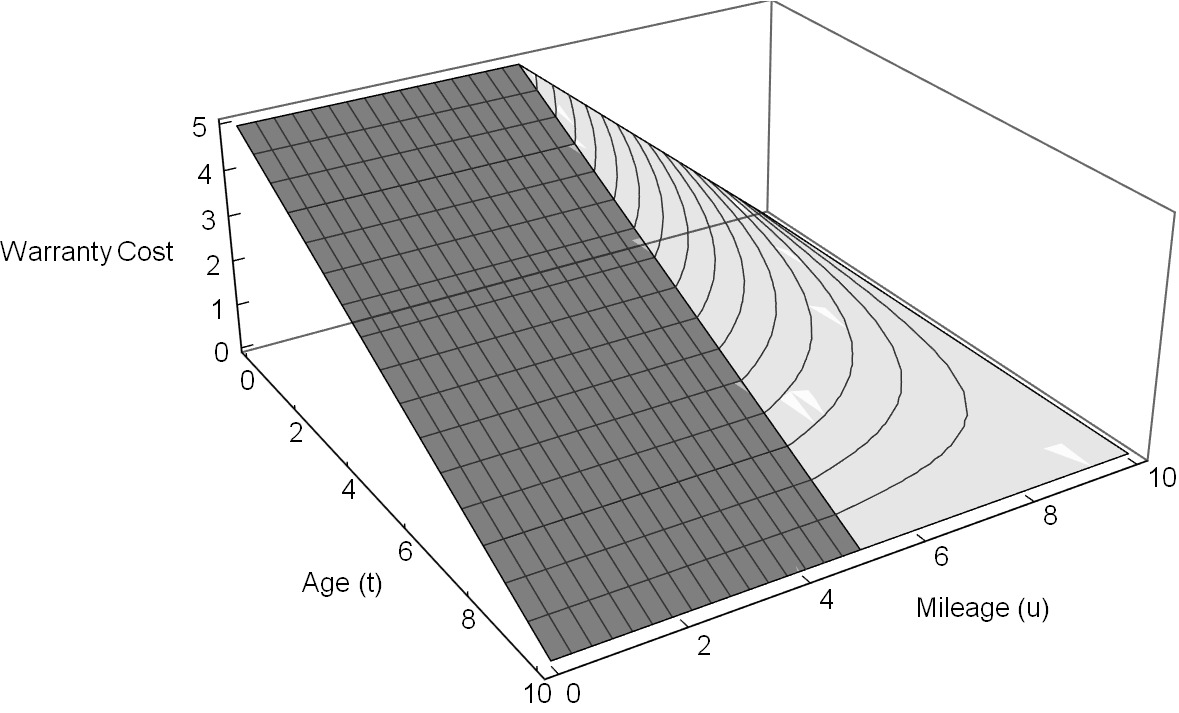}
    \caption{Warranty cost with age $t$ and mileage $u$ corresponding to case-VI}
    \label{Case-VI}
\end{figure}
    \item [Case-VII: ] $CW\times FRW$
\begin{align*}
    C_{CW\times FRW}^{T\times U}(t,u)=\begin{dcases}S & \text { if } 0\leq t\leq t_{w_1}, ~~0\leq u\leq u_{w}\\
    S\left(\frac{t_{w_2}-t}{t_{w_2}-t_{w_1}}\right)&\text{ if } t_{w_1}\leq t\leq t_{w_2}, ~~0\leq u\leq u_{w}
    \end{dcases}
\end{align*}
\begin{figure}[hbt!]
    \centering
    \includegraphics[scale=0.25]{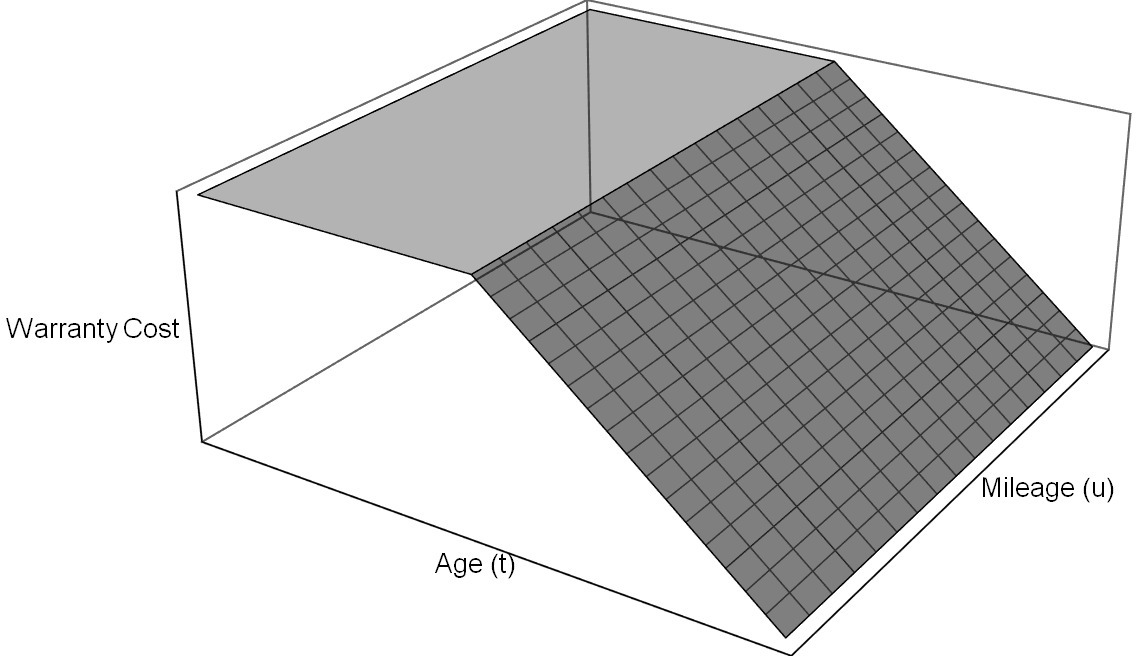}
    \caption{Warranty cost with age $t$ and mileage $u$ corresponding to case-VII}
    \label{Case-VII}
\end{figure}
    \item [Case-VIII: ] $CW\times PRW$
\begin{align*}
    C_{CW\times PRW}^{T\times U}(t,u)=\begin{dcases}S\left(1-\frac{u}{u_w}\right)& \text { if } 0\leq t\leq t_{w_1}, ~~0\leq u\leq u_{w}\\
    S\left(\frac{t_{w_2}-t}{t_{w_2}-t_{w_1}}\right)\left(1-\frac{u}{u_w}\right)&\text{ if } t_{w_1}\leq t\leq t_{w_2}, ~~0\leq u\leq u_{w}
    \end{dcases}
\end{align*}
\begin{figure}[hbt!]
    \centering
    \includegraphics[scale=0.25]{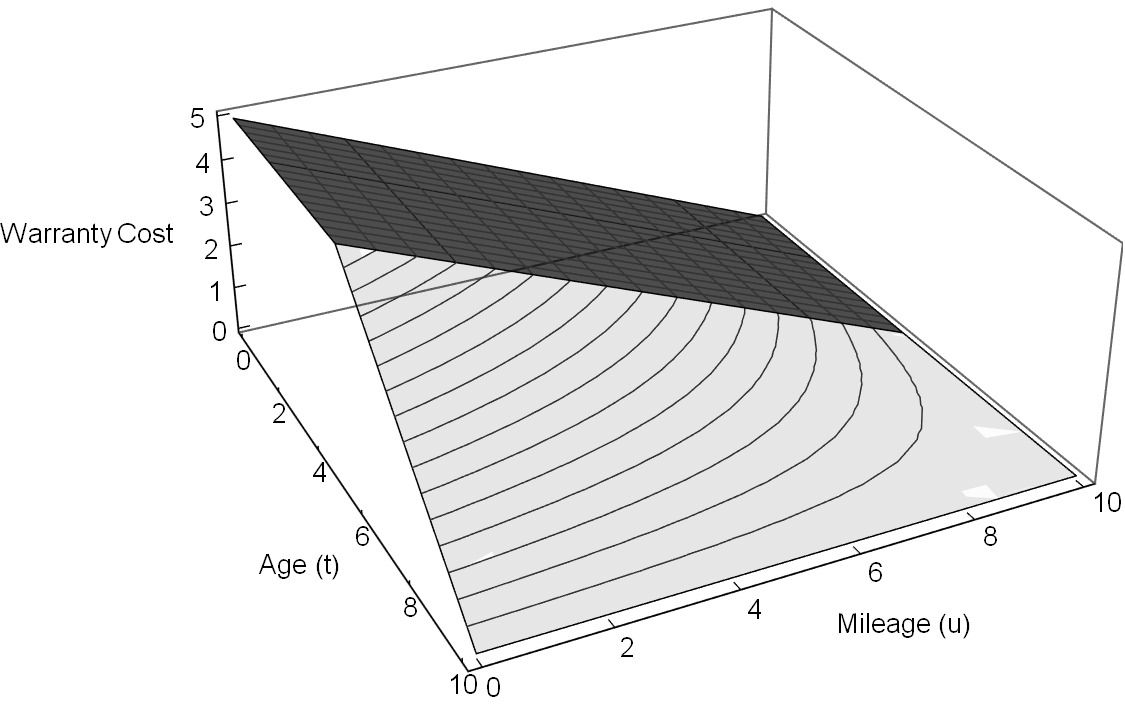}
    \caption{Warranty cost with age $t$ and mileage $u$ corresponding to case-VIII}
    \label{Case-VIII}
\end{figure}
    \item [Case-IX: ] $CW\times CW$
\begin{align*}
    C_{CW\times CW}^{T\times U}(t,u)=\begin{dcases}S & \text { if } 0\leq t\leq t_{w_1}, ~~0\leq u\leq u_{w_1}\\
    S\left(\frac{t_{w_2}-t}{t_{w_2}-t_{w_1}}\right)&\text{ if } t_{w_1}\leq t\leq t_{w_2}, ~~0\leq u\leq u_{w_1}\\
    S\left(\frac{u_{w_2}-u}{u_{w_2}-u_{w_1}}\right)&\text{ if } 0\leq t\leq t_{w_1}, ~~u_{w_1}\leq u\leq u_{w_2}\\
  S  \left(\frac{t_{w_2}-t}{t_{w_2}-t_{w_1}}\right)\left(\frac{u_{w_2}-u}{u_{w_2}-u_{w_1}}\right)&\text{ if } t_{w_1}\leq t\leq t_{w_2}, ~~u_{w_1}\leq u\leq u_{w_2}
    \end{dcases}
\end{align*}
\begin{figure}[hbt!]
    \centering
 \includegraphics[scale=0.2]{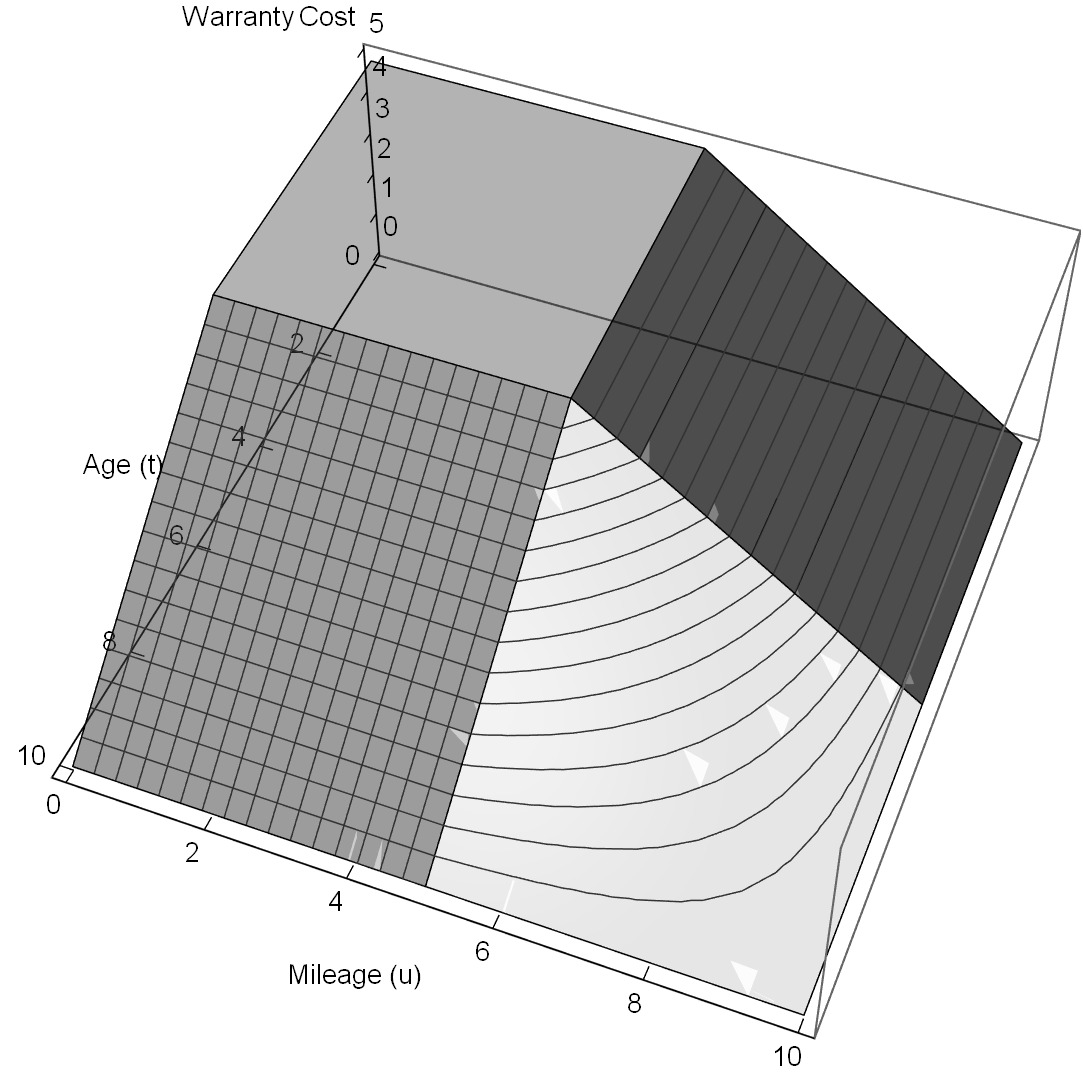} ~~~\includegraphics[scale=0.2]{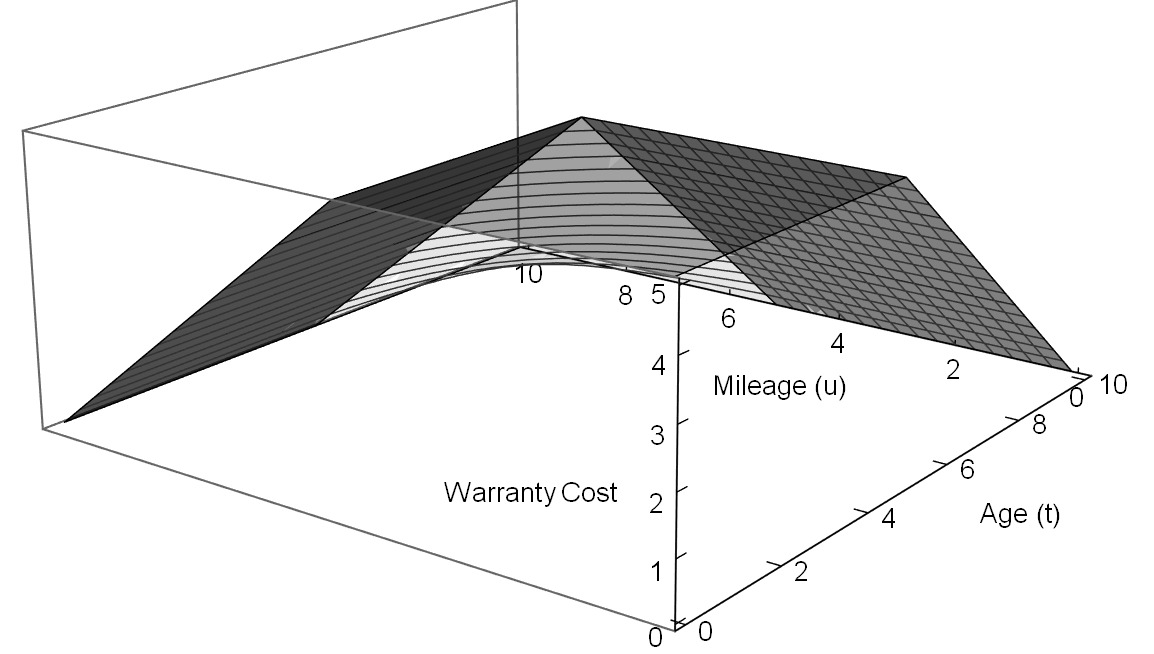}
    \caption{Warranty cost with age $t$ and mileage $u$ corresponding to case-XI}
    \label{Case-XI}
\end{figure}

\end{enumerate}

   It is noted from the scenario  \( CW \times CW \) that  we can derive all the remaining eight policies by considering special cases. Specifically, by taking \( x_{w_1} = x_{w_2} \), \( CW \) reduces to the \( FRW \) policy, and when \( x_{w_1} = 0 \), \( CW \) reduces to the \( PRW \) policy. \citet{wang2018two} proposed the $FRW$, $PRW$ and $CW$ policy  with the real-world warranty programs under rectangular region. According to the defination of $FRW$, $PRW$ and $CW$ policy in  \citet{wang2018two}, the Case- I is considered as $FRW$ policy, Case-II,IV,V,VI,VIII are PRW policy and remaining others consider as CW policy. 
   \section{Cost functions}
   In this article, two cost functions, such as the economic benefit function and warranty cost function, are considered, which were proposed by \citet{sen2022determination,Christen_2006}. In the subsequent sections, we have discussed those cost functions.
\subsection{Economic Benefit Function}
A warranty boosts product sales by assuring the buyer of compensation in case the product malfunctions. With a warranty in place, the manufacturer may experience an increase in sales volume, resulting in a monetary benefit. It would be unrealistic to assume that the benefit increases indefinitely as the warranty period increases. Given that a combined FRW-PRW policy includes two stages of warranty, we assume that the benefit grows with the average of the two warranty lengths for each dimension. Furthermore, as age $t$ and usages $u$ tend to infinity, each individual benefit function approaches 1, and consequently, their product is bounded by 1. Therefore, we consider the benefit function to be bounded as the warranty lengths approach infinity. We propose the economic benefit function for two dimensional warranty as the product of two exponential decay function with bounded above by $A_1M$ 

\begin{eqnarray}\label{economic_benefit1}
    B(t_{w_1},t_{w_2},u_{w_1},u_{w_2})=A_1M B^{(T)}(t_{w_1}, t_{w_2}) \times B^{(U)}(u_{w_1}, u_{w_2}).
\end{eqnarray}
where the economic benefit for time/age represents as 
    \[
    B^{(T)}(t_{w_1}, t_{w_2}) = 1 - \exp\left\{-A_2 \left(\frac{t_{w_1} + t_{w_2}}{2}\right)\right\}, 
    \]
    as $t_{w_1}, t_{w_2} \to 0$, $B^{(T)} \to 0$ and $t_{w_1}, t_{w_2} \to \infty$, $B^{(T)}  \to 1$, thus $0 \leq B^{(T)}(t_{w_1}, t_{w_2}) \leq 1$, and
    the economic benefit of usage represents
    $$ B^{(U)}\left(u_{w_1}, u_{w_2}\right) = 1 - \exp\left\{-A_3 \left(\frac{u_{w_1} + u_{w_2}}{2}\right)\right\}$$
     as $u_{w_1}, u_{w_2} \to 0$, $B^{(U)} \to 0$ and  $u_{w_1}, u_{w_2} \to \infty$, $B^{(U)} \to 1$, thus $0 \leq  B^{(U)}(u_{w_1}, u_{w_2}) \leq 1$.
Hence $0 \leq B(t_{w_1}, t_{w_2}, u_{w_1}, u_{w_2}) \leq A_1 M$. \\

The parameters \( A_1 \) represent the manufacturer’s profit per product, \( M \) denotes the potential number of products that could be sold under this warranty policy. for the scales $t$ and $u$, parameters \( A_2, A_3 \), respectively, controls the rate at which the benefit increases. The values of \( A_2, A_3 \) can be uniquely determined based on the ratio of two special cases ($w_1 = 0$ and $w_1 = w_2$)  of economic benefit function (\ref{economic_benefit1}) in the combined FRW-PRW policy for both time and usage scale. Here the ratio means that the percentage of benefit remains when the manufacturer changes the warranty policy from FRW to PRW. Consider  standard market warranties $t_w$ and $u_w$, respectively under FRW policy for both the scales.  To derive $A_2$, we construct the ratio $h(A_2)$ by evaluating (\ref{economic_benefit1}) at $t_{w_1} = 0$ (CW become PRW) and $t_{w_1} = t_{w_2}$ (CW become PRW), corresponding to a change in the t-axis policy from FRW to PRW, while keeping the u-axis policy fixed as FRW. The ratio is 
\begin{align}\label{t1}
    h_1(A_2) = \frac{B(0,t_w,u_w,u_w)}{B(t_w,t_w,u_w,u_w)}=\frac{1-\exp\left(-\frac{A_2t_w}{2}\right)}{1-\exp\left(-A_2t_w\right)}.
\end{align}

It is easily shown that $h_1$ is increasing function with $h_1(0+)=1/2$ and $h_1(\infty)=1$. By taking a value $q_1^*$ between 0.5 and 1, $A_2$ can be derived by solving the non linear equation $h(A_2) = q^*$. Similarly, for the parameter $A_3$ the ratio is  
\begin{align}\label{t2}
   h_2(A_3) =  \frac{B(t_w,t_w,0,u_w)}{B(t_w,t_w,u_w,u_w)}=\frac{1-\exp\left(-\frac{A_3u_w}{2}\right)}{1-\exp\left(-A_3u_w\right)}.
\end{align}
It may be noted that here u-axis policy changes from FRW to PRW, while keeping the t-axis policy fixed. The ratio means that the percentage of benefit remains when the manufacturer changes the warranty policy from FRW to PRW with respect to mileage $u$. 
It is easily shown that $h_2$ is increasing function with $h_2(0+)=1/2$ and $h_2(\infty)=1$. Hence by solving the non linear equation $h(A_3) = q_2^*$, we can derived $A_3$. Also One can consider the combined ratio
\begin{align}\label{t3}
     \frac{B(0,t_w,0,u_w)}{B(t_w,t_w,u_w,u_w)}=\frac{1-\exp\left(-\frac{A_2t_w}{2}\right)}{1-\exp\left(-A_2t_w\right)}\times\frac{1-\exp\left(-\frac{A_3u_w}{2}\right)}{1-\exp\left(-A_3u_w\right)}=h_1(A_2)\times h_2(A_3),
\end{align}
which means that the percentage of benefit remains when the manufacturer changes the warranty policy from FRW to PRW with respect to both variable age $t$ and mileage $u$. From the above two conditions, we get the percentage value.
\subsection{Warranty Cost per Unit Sale}
Warranty cost represents the direct expense to the manufacturer for reimbursing products that fail during the warranty period. Let the warranty cost be denoted by \(W(t_{w_1},t_{w_2},u_{w_1},u_{w_2}) \). It is defined as

\begin{eqnarray*}
W(t, u, t_{w_1},t_{w_2},u_{w_1},u_{w_2}) & = & \{ \mbox{the expected
number of items that fail under the warranty period} \}\\
&& ~~~\times  ~\{\mbox{cost of reimbursing an item } C_{P_1\times P_2}^{T\times U}(t,u)\}
\end{eqnarray*}
  
\noindent The warranty cost function can be written as
\allowdisplaybreaks\begin{align*}
W(t, u, t_{w_1},t_{w_2},u_{w_1},u_{w_2})
&=MF(t_{w_1},u_{w_1})SI_{[0,t_{w_1})\times[0,u_{w_1})}(t,u)\\
&+M[F(t_{w_2},u_{w_1})-F(t_{w_1},u_{w_1})]S\frac{t_{w_2}-t}{t_{w_2}-t_{w_1}}I_{[t_{w_1},t_{w_2})\times[0,u_{w_1})}(t,u)\\
    &+M[F(t_{w_1},u_{w_2})-F(t_{w_1},u_{w_1})]S\frac{u_{w_2}-u}{u_{w_2}-u_{w_1}}I_{[0,t_{w_1})\times[u_{w_1},u_{w_2})}(t,u)\\
    &+M[F(t_{w_2},u_{w_2})+F(t_{w_1},u_{w_1})-F(t_{w_2},u_{w_1})-F(t_{w_1},u_{w_2})] \\ 
    & ~~~~~~~~~~~~\times S\frac{t_{w_2}-t}{t_{w_2}-t_{w_1}}\frac{u_{w_2}-u}{u_{w_2}-u_{w_1}}I_{[t_{w_1},t_{w_2})[u_{w_1},u_{w_2})}(t,u),\\
\end{align*}

where:
\begin{eqnarray*}
    P(0\leq T<t_{w_1},0\leq U<u_{w_1}) & = &  F(t_{w_1},u_{w_1}) \\
    P(t_{w_1}\leq T<t_{w_2},0 \leq U< u_{w_1}) & = & F(t_{w_2},u_{w_1})-F(t_{w_1},u_{w_1})\\
    P(0\leq T<t_{w_1},u_{w_1}\leq U<u_{w_2}) &=& F(t_{w_1},u_{w_2})-F(t_{w_1},u_{w_1})\\
    P(t_{w_1}\leq T<t_{w_2}, u_{w_1}\leq U< u_{w_2}) & = & F(t_{w_2},u_{w_2})+F(t_{w_1},u_{w_1})-F(t_{w_2},u_{w_1})-F(t_{w_1},u_{w_2}),
\end{eqnarray*}

and 
\begin{itemize}
    \item $\mathbb{I}_{[a,b) \times [c,d)}(t,u)$ is the indicator function that equals 1 when $(t,u) \in [a,b) \times [c,d)$ and 0 otherwise,
    \item $S$ is the cost per claim,
    \item $M$ represents the potential number of products that could be sold under this warranty policy.
\end{itemize}



\section{Optimal Warranty Region}
The utility function defined in this paper is composed of two functions: the economic benefit function, and the warranty cost function. The resulting utility function is then expressed as:
\begin{eqnarray}
    U(t, u, t_{w_1},t_{w_2},u_{w_1},u_{w_2}) =  B(t_{w_1},t_{w_2},u_{w_1},u_{w_2}) -W(t, u, t_{w_1},t_{w_2},u_{w_1},u_{w_2}). \label{utility}
\end{eqnarray}

By taking the expectation on both sides of (\ref{utility}), we can derive the expected utility function, which will determine the optimal warranty region. The expression is given below
\begin{eqnarray}
    E[U(t, u, t_{w_1}, t_{w_2}, u_{w_1}, u_{w_2})] & = & B(t_{w_1},t_{w_2},u_{w_1},u_{w_2}) -E[W(t, u, t_{w_1},t_{w_2},u_{w_1},u_{w_2})] \nonumber \\    
    &=& B(t_{w_1},t_{w_2},u_{w_1},u_{w_2}) - \int_{0}^\infty W(t, u, t_{w_1}, t_{w_2}, u_{w_1}, u_{w_2})~ f(t, u)~ dt~ du. \label{expected_utility}
\end{eqnarray}

\noindent The expected warranty cost derived is given below
\begin{align*}
E[W(t, u, t_{w_1},t_{w_2},u_{w_1},u_{w_2})] = &~~~~ M F(t_{w_1},u_{w_1}) S \times W^{(1)} \\
&+ M  \big[F(t_{w_2},u_{w_1}) - F(t_{w_1},u_{w_1})\big] ~S \times  W^{(2)} \\
& + M \big[F(t_{w_1},u_{w_2}) - F(t_{w_1},u_{w_1})\big] ~S \times W^{(3)} \\
& + M  \big[F(t_{w_2},u_{w_2}) + F(t_{w_1},u_{w_1}) - F(t_{w_2},u_{w_1}) - F(t_{w_1},u_{w_2})\big] ~S \times W^{(4)},
\end{align*}
where 
\begin{eqnarray*}
    W^{(1)} &=& \int_{0}^{t_{w_1}} \int_{0}^{u_{w_1}} f(t,u) \, dt \, du = F(t_{w_1}, u_{w_1}) \\
   W^{(2)} & = & \int_{t_{w_1}}^{t_{w_2}} \int_{0}^{u_{w_1}} \frac{t_{w_2} - t}{t_{w_2} - t_{w_1}} f(t,u) \, du \, dt \\
   W^{(3)} & = & \int_{0}^{t_{w_1}} \int_{u_{w_1}}^{u_{w_2}} \frac{u_{w_2} - u}{u_{w_2} - u_{w_1}} f(t,u) \, du \, dt \\
   W^{(4)} & = & \int_{t_{w_1}}^{t_{w_2}} \int_{u_{w_1}}^{u_{w_2}} \frac{(t_{w_2} - t)(u_{w_2} - u)}{(t_{w_2} - t_{w_1})(u_{w_2} - u_{w_1})} f(t,u) \, du \, dt.\\
\end{eqnarray*}

The optimal warranty region $(t_{w_1}^*, t_{w_2}^*, u_{w_1}^*, u_{w_2}^*)$ can be obtained by maximizing the expected value of the utility function given in (\ref{expected_utility}). In other words, 
\begin{eqnarray}
   t_{w_1}^*, t_{w_2}^*, u_{w_1}^*, u_{w_2}^* = \arg \left( \max_{w_1 \leq w_2;  u_1 \leq u_2 }~ E[U(t, u, t_{w_1}, t_{w_2}, u_{w_1}, u_{w_2})] \right). \label{opt_utility}
\end{eqnarray}
Algorithm 1 is proposed to compute the optimum solution for the above minimization problem (\ref{opt_utility}).


\begin{algorithm}[hbt!]
\caption{}
\textbf{Step 1.} Set the values of the parameters $\boldsymbol{\theta}$, $S$, $A_1$, $t_w$ and $u_w$.

\textbf{Step 2.} Choose the values $q_1^*$ and $q_2^*$ from  $[1/2,~ 1)$. 

\textbf{Step 3.} To determine $A_2$ and $A_3$, solve the non linear equations $h(A_2) = q^*$ and $h(A_3) = q^*$. 

\textbf{Step 3.} Consider two variables $k_1, k_2$. We set $t_{w_2}=t_{w_1}+k_1$ and $u_{w_2}=u_{w_1}+k_2$. 

\textbf{Step 4:} Solve a constrained non linear Minimization problem in  (\ref{opt_utility}) with respect to $(t_{w_1},k_1,u_{w_1},k_2)$. 

\textbf{Step 5:} Let $k_1^*$ and $k_2^*$ be the optimal values of $k_1$ and $k_2$, and let $t_{w_1}^*$ and $u_{w_1}^*$ be the optimal values of $t_{w_1}$ and $u_{w_1}$, respectively. Then, $t_{w_2}^* = t_{w_1}^* + k_1^*$ and $u_{w_2}^* = u_{w_1}^* + k_2^*$ be  the optimal values of $t_{w_2}$ and $u_{w_2}$.

\end{algorithm}


    
    


\section{Real Life Data and Numerical Example}
\begin{figure}
    \centering
    \includegraphics[scale=0.7]{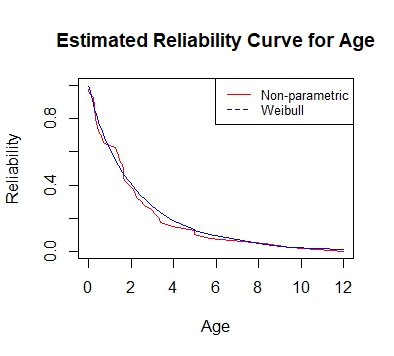}
\includegraphics[scale=0.7]{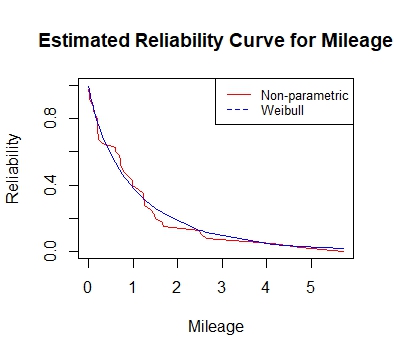}
    \caption{Reliability curve }
    \label{fig}
\end{figure}

The proposed methodology for determining the optimal warranty region is demonstrated using bivariate real-life failure data obtained from the maintenance records of $n = 40$ locomotive traction motors, given in Table \ref{data1}. The original data was presented in \cite{eliashberg1997calculating}. We first fit this data to various copula-based models and observed from Table \ref{fit_copula} that among them, the Gumbel copula is the best-fitting model with a significantly higher $p$-value.  Additionally, we fit the age and mileage data for each scale separately with Weibull distributions to validate the goodness of fit both graphically and through hypothesis testing.
We get the MLEs for the age data as $\widehat{\lambda}_T=2.243$ and $\widehat{\eta}_T=0.897$ and for mileage data as $\widehat{\lambda}_U=1.046$ and $\widehat{\eta}_U=0.830$. To verify whether the age and mileage data are well described by the marginal Weibull distributions, the parametric reliability curve using the MLEs and the non-parametric reliability curve using the Kaplan-Meier estimator for both age and mileage are shown in Figure \ref{fig}.
Solid lines show Weibull fits (MLE), dashed lines represent Kaplan-Meier estimates.  From Figure \ref{fig}, it is observed that both age and mileage are well fitted to the Weibull distribution, as both the curves close to each other. The Anderson-Darling goodness-of-fit test is also applied to the fitted distributions. The results of the test at a 5\% significance level indicate that the Weibull distribution fits the age data best ($p$-value = 0.8961) and also fits the mileage data well (p-value = 0.7896).

\begin{table}[hbt!]\centering
\caption{Bivariate failure Data}
\label{data1}
\begin{tabular}{c c c| c c c| c c c| c c c}
\hline
No. & Age & Mileage & No. & Age & Mileage & No. & Age & Mileage & No. & Age & Mileage \\
\hline
1 & 1.66 & 0.9766 &11 & 3.35 & 1.3827 & 21 & 1.28 & 0.5922 &31 & 0.01 & 0.0028 \\
2 & 0.35 & 0.2041 &12 & 1.64 & 0.5992& 22 & 0.31 & 0.1974 &32 & 0.27 & 0.0095  \\
3 & 2.49 & 1.2392 & 13 & 1.45 & 0.6925&23 & 0.65 & 0.2030 &33 & 2.95 & 1.2600  \\
4 & 1.90 & 0.9889 &14 & 1.70 & 0.7078 &24 & 2.21 & 1.2532 &  34 & 1.40 & 0.8067  \\
5 & 0.27 & 0.0974 &15 & 1.40 & 0.7553 & 25 & 3.16 & 1.4796 &  35 & 8.27 & 4.1425  \\
6 & 0.41 & 0.1594 & 16 & 4.98 & 2.5014 & 26 & 0.22 & 0.0979 & 36 & 0.02 & 0.0105 \\
7 & 0.59 & 0.2128 &17 & 5.71 & 2.5380 & 27 & 2.61 & 1.5062 & 37 & 2.09 & 1.2302\\
8 & 0.75 & 0.2158 &18 & 4.99 & 2.6433 & 28 & 0.32 & 0.2062 &  38 & 0.29 & 0.0447 \\
9 & 2.23 & 1.1187 & 19 & 3.40 & 1.6494 & 29 & 3.97 & 1.6888& 39 & 1.66 & 0.9766  \\
10 & 9.52 & 4.7660 & 20 & 1.60 & 0.7162 &30 & 0.48 & 0.3099& 40 & 12.00 & 5.7304   \\
\hline
\end{tabular}
\end{table}

\begin{table}[hbt]
\caption{p-value for different copulas}
    \label{tab:my_label}
    \centering
    \begin{tabular}{c| c c c c c}
    \hline
    \label{fit_copula}
    Copula& Normal & Clayton&Gumbel&Frank&Joe\\
    \hline
      p-value  &  0.3312& 0.0055 &0.6658 &0.1933&0.6558\\
      \hline
    \end{tabular}
\end{table}


The maximum likelihood estimators (MLEs) $\widehat{\boldsymbol{\psi}}=(\widehat{\lambda}_T$, $\widehat{\lambda}_U$, $\widehat{\eta}_T$, $\widehat{\eta}_U$, $\widehat{\theta}$) of $\boldsymbol{\psi}=(\lambda_T,\lambda_U,\eta_T,\eta_U,\theta)$, can be obtained by maximizing the log-likelihood function (\ref{log_like}). The estimated MLE for the real data is obtained as $\widehat{\psi}=(2.1807, 1.0398, 0.9132, 0.8518, 6.5937)$. 

For example, assume that the sales price of the product is \(S = 700\) and the production cost is \(C = 500\). As a result, the per unit profit is \(A_1 = 200\). Under the FRW policy, the manufacturer provides a standard warranty for both scales (age and mileage), which corresponds to the $0.1$th quantile of each marginal distribution, i.e., $t_w=0.1855$ and $u_w=0.0741$. Since the manufacturers are interested in the combined FRW-PRW policy, shifting from the FRW to the PRW policy results in a percentage of benefit, denoted as \( q_1^*, q_2^* \) for the t and u axes. As discussed in Section 3.1, we have chosen the percentage of benefit \( q_1^* = q_2^* = 0.75 \), which lies between 0.5 and 1, for both scales.  To determine the control parameters \( A_2 \) and \( A_3 \) in the economic benefit function, we propose solving the two nonlinear equations \( h(A_2) = q_1^* \) and \( h(A_3) = q_2^* \), respectively. After solving these two equations, we obtain the unique solutions  $A_2=11.844$, $A_3=29.665$. 


The optimal warranty region for all possible combinations of warranty policies is computed and shown in Tables \ref{cost500}, \ref{cost700},  and \ref{cost900}, based on sales prices of 500, 700, and 900, respectively. From Table \ref{cost700}, it is observed that if both the scales are in combined policy i.e., \( CW \times CW \) we achieve significantly higher utility, with a comparatively larger warranty length for both age and mileage. It is also observed from the table that if we consider either the FRW policy solely (i.e., \( FRW \times FRW \)) or the PRW policy solely \(i.e.,  PRW \times PRW \)  for both scales, the achieved utility is significantly lower compared to other policies, with \( PRW \times PRW \) ranking last in the table. Similar patterns are also observed in the other two tables, Table \ref{cost500} and Table \ref{cost900}. It is observed from the Tables that when selling price increases then expected utility increases and warranty region become wider as expected.

\begin{table}[hbt!]
    \centering
    \caption{Two-dimensional optimal warranty design for different cases when $S=500$}
    \label{cost500}
    \begin{tabular}{lccccS[table-format=3.6]}
    \toprule
    Policy & \multicolumn{4}{c}{Warranty Design} & {Utility} \\
    \cmidrule(lr){2-5}
     & {$t^*_{w_1}$} & {$t^*_{w_2}$} & {$u^*_{w_1}$} & {$u^*_{w_2}$} & \\
    \midrule
    CW × CW &0.2742& 0.5292& 0.0496 &0.3914 & 191.9923 \\
     CW × PRW &0.1964& 0.5478&0.0000&  0.4072 & 188.8317 \\
    CW × FRW &  0.1558& 1.0992& 0.1508 & 0.1508&190.5998 \\
    PRW × CW & 0.0000 &  1.0816 &0.0774 &0.2219& 189.2955 \\
    PRW × PRW &0.0000& 0.8187&0.0000& 0.2627 & 179.9070 \\
     PRW × FRW &0.0000&1.0858& 0.1334&0.1334 & 185.0547 \\
    FRW × CW &0.3716& 0.3716& 0.0620& 0.4076 &  190.1626 \\
    FRW × PRW & 0.3269&0.3269&0.0000& 0.4237 & 184.4741 \\
    FRW × FRW & 0.3967&0.3967& 0.1469&0.1469 & 184.5439 \\
    
    \bottomrule
    \end{tabular}
\end{table}

\begin{table}[hbt!]
    \centering
    \renewcommand{\arraystretch}{1}  
    \caption{Two-dimensional optimal warranty design for different cases when $S=700$}
    \label{cost700}
    \begin{tabular}{lccccS[table-format=3.6]}
    \toprule
    Policy & \multicolumn{4}{c}{Warranty Design} & {Utility} \\
    \cmidrule(lr){2-5}
     & {$~~~~t^*_{w_1}~~~~$} & {$t^*_{w_2}$} & {$u^*_{w_1}$} & {$u^*_{w_2}$} & \\
    \midrule
    CW × CW & 0.2578 & 0.4884 & 0.0470 & 0.3854 & 189.7210 \\
     CW × PRW & 0.1844 & 0.5037 & 0.0000 & 0.3953 & 185.7155 \\
    CW × FRW & 0.1478 & 1.0788 & 0.1402 & 0.1402 & 187.9544 \\
    PRW × CW & 0.0000 & 1.0430 & 0.0725 & 0.2053 & 186.2616 \\
    PRW × PRW & 0.0000 & 0.7834 & 0.0000 & 0.2397 & 174.6750 \\
    PRW × FRW & 0.0000 & 1.0503 & 0.1230 & 0.1230 & 180.9913 \\
    FRW × CW & 0.3447 & 0.3447 & 0.0583 & 0.4037 & 187.4519 \\
     FRW × PRW & 0.3008 & 0.3008 & 0.0000 & 0.4120 & 180.3577 \\
    FRW × FRW & 0.3708 & 0.3708 & 0.1373 & 0.1373 & 180.3717 \\
    \bottomrule
     \end{tabular}
\end{table}

\begin{table}[hbt!]
    \centering
    \caption{Two-dimensional optimal warranty design for different cases when $S=900$}
    \label{cost900}
    \begin{tabular}{lccccS[table-format=3.6]}
    \toprule
    Policy & \multicolumn{4}{c}{Warranty Design} & {Utility} \\
    \cmidrule(lr){2-5}
     & {$t^*_{w_1}$} & {$t^*_{w_2}$} & {$u^*_{w_1}$} & {$u^*_{w_2}$} & \\
    \midrule
    CW × CW &0.2454& 0.4587 &0.0451& 0.3823& 187.658 \\
     CW × PRW & 0.1753 & 0.4713 & 0.0000 & 0.3875 & 182.8983 \\
    CW × FRW & 0.1417 & 1.0662 & 0.1324& 0.1324 & 185.5544 \\
     PRW × CW & 0.0000 & 1.0161 & 0.0688 & 0.1930 & 183.5042 \\
      PRW × PRW & 0.0000 & 0.7636 & 0.0000 & 0.2220 & 169.5915 \\
       PRW × FRW & 0.0000 & 1.0258 & 0.1153 & 0.1153 & 177.3422 \\ 
    FRW × CW & 0.3250 & 0.3250 & 0.0556 & 0.4021 & 185.0077 \\ 
    FRW × PRW & 0.2816 & 0.2816 & 0.0000 & 0.4043 & 176.6866 \\
    FRW × FRW & 0.3515 & 0.3515 & 0.1302 & 0.1302 & 176.6192 \\    
    \bottomrule
    \end{tabular}
\end{table}
\section{Conclusion}
In this work, we propose a two-dimensional warranty cost function by considering all possible scenarios under a combined policy. A bivariate Gumbel copula is utilized to model positively correlated failure time data, where the Weibull distribution is the best-fitted model. The numerical studies indicate that a combined policy considering both dimensions, i.e., age and usage, leads to a significantly higher utility than any other scenario. On the other hand, when the manufacturer provides a prorata Warranty (PRW) for both age and usage, the utility is considerably lower. 

The work can be extended to $n$-dimensional warranty. If $X_1,\ldots,X_n$ are random variables of $n$ measurable quantities to the failure of the product, the cost of reimbursement of a product can be written as
\begin{align*}
    C^{(X_1\times\ldots\times X_n)}_{W_1\times\ldots\times W_n}(x_1,\ldots,x_n)=\frac{1}{S^{n-1}}\prod_{i=1}^nC^{X_i}_{W_i}(x_i)
\end{align*}
Also, in this work, we use linear pro-rated compensation for each individual scales. However, the work can be extended to use non-linear function for pro-rated compensation for the individual scales (see \citet{sen2022determination}).
\bibliographystyle{apalike}
\bibliography{BaysWarranty}

\begin{thebibliography}{}

\bibitem[Blischke and Murthy, 1992]{blischke1992product}
Blischke, W.~R. and Murthy, D. (1992).
\newblock {Product warranty management—I: A taxonomy for warranty policies}.
\newblock {\em European journal of operational research}, 62(2):127--148.

\bibitem[Dai et~al., 2017]{dai2017field}
Dai, A., He, Z., Liu, Z., Yang, D., and He, S. (2017).
\newblock Field reliability modeling based on two-dimensional warranty data with censoring times.
\newblock {\em Quality Engineering}, 29(3):468--483.

\bibitem[Eliashberg et~al., 1997]{eliashberg1997calculating}
Eliashberg, J., Singpurwalla, N.~D., and Wilson, S.~P. (1997).
\newblock Calculating the reserve for a time and usage indexed warranty.
\newblock {\em Management Science}, 43(7):966--975.

\bibitem[Gupta et~al., 2014]{gupta2014warranty}
Gupta, S.~K., De, S., and Chatterjee, A. (2014).
\newblock Warranty forecasting from incomplete two-dimensional warranty data.
\newblock {\em Reliability Engineering \& System Safety}, 126:1--13.

\bibitem[Guti{\'e}rrez-Pulido et~al., 2006]{Christen_2006}
Guti{\'e}rrez-Pulido, H., Aguirre-Torres, V., and Christen, J.~A. (2006).
\newblock {A Bayesian approach for the determination of warranty length}.
\newblock {\em Journal of Quality Technology}, {\bf{38}}:180 -- 189.

\bibitem[Jung and Bai, 2007]{jung2007analysis}
Jung, M. and Bai, D.~S. (2007).
\newblock Analysis of field data under two-dimensional warranty.
\newblock {\em Reliability Engineering \& System Safety}, 92(2):135--143.

\bibitem[Lawless et~al., 1995]{lawless1995methods}
Lawless, J., Hu, J., and Cao, J. (1995).
\newblock Methods for the estimation of failure distributions and rates from automobile warranty data.
\newblock {\em Lifetime Data Analysis}, 1:227--240.

\bibitem[Manna et~al., 2006]{manna2006optimal}
Manna, D., Pal, S., and Sinha, S. (2006).
\newblock Optimal determination of warranty region for 2d policy: A customers' perspective.
\newblock {\em Computers \& Industrial Engineering}, 50(1-2):161--174.

\bibitem[Manna et~al., 2007]{manna2007use}
Manna, D., Pal, S., and Sinha, S. (2007).
\newblock A use-rate based failure model for two-dimensional warranty.
\newblock {\em Computers \& Industrial Engineering}, 52(2):229--240.

\bibitem[Manna et~al., 2008]{manna2008note}
Manna, D., Pal, S., and Sinha, S. (2008).
\newblock A note on calculating cost of two-dimensional warranty policy.
\newblock {\em Computers \& Industrial Engineering}, 54(4):1071--1077.

\bibitem[Menke, 1969]{menke1969determination}
Menke, W.~W. (1969).
\newblock {Determination of warranty reserves}.
\newblock {\em Management Science}, 15(10):B--542.

\bibitem[Murthy and Blischke, 2006]{blischke_2006}
Murthy, D. N.~P. and Blischke, W.~R. (2006).
\newblock {\em Warranty management and product manufacture}.
\newblock Springer Science \& Business Media.

\bibitem[Nelsen, 2006]{nelsen2006introduction}
Nelsen, R.~B. (2006).
\newblock {\em An introduction to copulas}.
\newblock Springer.

\bibitem[Sen et~al., 2022]{sen2022determination}
Sen, T., Bhattacharya, R., Pradhan, B., and Tripathi, Y.~M. (2022).
\newblock {Determination of Bayesian optimal warranty length under Type-II unified hybrid censoring scheme}.
\newblock {\em Quality Technology \& Quantitative Management}, 19(1):35--49.

\bibitem[Sklar, 1959]{sklar1959fonctions}
Sklar, M. (1959).
\newblock Fonctions de r{\'e}partition {\`a} n dimensions et leurs marges.
\newblock In {\em Annales de l'ISUP}, volume~8, pages 229--231.

\bibitem[Thomas, 1983]{thomas1983optimum}
Thomas, M.~U. (1983).
\newblock {Optimum warranty policies for nonreparable items}.
\newblock {\em IEEE transactions on reliability}, 32(3):282--288.

\bibitem[Wang and Xie, 2018]{wang2018two}
Wang, X. and Xie, W. (2018).
\newblock {Two-dimensional warranty: A literature review}.
\newblock {\em Proceedings of the Institution of Mechanical Engineers, Part O: Journal of Risk and Reliability}, 232(3):284--307.

\bibitem[Wu et~al., 2007]{wu2007optimal}
Wu, C.-C., Chou, C.-Y., and Huang, C. (2007).
\newblock Optimal burn-in time and warranty length under fully renewing combination free replacement and pro-rata warranty.
\newblock {\em Reliability Engineering \& System Safety}, 92(7):914--920.

\bibitem[Wu and Huang, 2010]{wu_2010}
Wu, S.~J. and Huang, S.~R. (2010).
\newblock {Optimal warranty length for a Rayleigh distributed product with progressive censoring}.
\newblock {\em IEEE Transactions on Reliability}, {\bf{59}}:661 -- 666.

\bibitem[Yang et~al., 2016]{yang2016warranty}
Yang, D., He, Z., and He, S. (2016).
\newblock Warranty claims forecasting based on a general imperfect repair model considering usage rate.
\newblock {\em Reliability Engineering \& System Safety}, 145:147--154.

\end{thebibliography}

\end{document}